\documentclass[12pt]{iopart}

\usepackage{iopams} 
\usepackage{graphicx}

\expandafter\let\csname equation*\endcsname\relax 
\expandafter\let\csname endequation*\endcsname\relax 
\usepackage{amsmath}
\usepackage{mathtools}

\begin{document}

\title{Matrix continued fraction solution to the relativistic spin-$0$ Feshbach-Villars equations }

\author{N.~C.~Brown, Z.~Papp and R.~Woodhouse }
\address{ Department of Physics and Astronomy,
California State University Long Beach, Long Beach, California, USA }

\date{\today}

\begin{abstract}
The Feshbach-Villars equations, like the Klein-Gordon equation, 
are relativistic quantum mechanical equations for spin-$0$ particles.
We write the  Feshbach-Villars equations into an integral equation form and 
solve them by applying the Coulomb-Sturmian potential separable expansion method. 
We consider bound-state problems in a Coulomb plus short range 
potential. The corresponding Feshbach-Villars Coulomb Green's operator is represented by a 
matrix continued fraction. 

\end{abstract}


\maketitle

\section{Introduction}
\label{intro}

It is well known that the basic relativistic  equation for a spin-$0$ particle is the Klein-Gordon equation. 
Probably, it had been formulated first by Schr\"odinger, but it gave wrong results for the hydrogen atom, 
so he abandoned it in
favor of the non-relativistic formalism.
The Klein-Gordon equation was suggested 
in 1926 by Klein \cite{klein1926quantum} and Gordon \cite{gordon1926zs} as well as by several other people \cite{fock,kudar,donder}. 
It can be derived from the relativistic energy-momentum relation using the correspondence principle. 
It has served as a basic relativistic equation for 
spin-$0$ particles in relativistic quantum mechanics and field theory.

The Klein-Gordon equation, however, contradicts some of the postulates of quantum mechanics. 
In the quantum mechanics it is postulated  that the system is 
completely determined by the wave function and the time evolution of the wave function is determined 
by the time-dependent
 Schr\"odinger equation \cite{zettili,shankar}. The Klein-Gordon
equation contains a second derivative in time, therefore to determine the system the wave function in 
not enough, we
need its time derivative as well. 
So, obviously, the Klein-Gordon equation is not a genuine quantum mechanical equation.

In 1958 Feshbach and Villars rewrote the Klein-Gordon equation into a Hamiltonian form
 \cite{Feshbach:1958wv}. They split
the Klein-Gordon wave function into two components and  for the components vector they
 arrived at a Schr\"odinger-like
equation with first order in time derivative. Although the Feshbach-Villars formalism appear in some advanced 
quantum mechanics books 
\cite{corinaldesi2015relativistic,davydov1976quantum,baym1969lectures,greiner1990relativistic,ni2002advanced,wachter2010relativistic,strange1998relativistic}, 
and they were utilized in gaining deeper insight into relativistic physics
of Klein paradox pair production, \cite{wagner2010exponential,haouat2005pair,wagner2010bosonic,bounames2007klein,bounames2001solution,chetouani2004solution,merad2000boundary,guettou2006pair}, in exotic atoms \cite{leon1981relativistic,fuda1980feshbach,friar1980feshbach}, used in theoretical consideraions \cite{wong2010klein,kowalenko1985response,sidharth2011negative,rizov1985relativistic}, study relativistic
 scattering \cite{khounfais2004scattering,khounfais2004scattering} and optics \cite{haghighat2003coherent} or demosntrate PT symmetry \cite{znojil2005solvable,znojil2004experiments,znojil2004pseudo,znojil2004relativistic}, they were hardly used
as a computational tool.
The equations look like ordinary coupled differential
equations, but the components are coupled by the kinetic energy operator, which makes them very hard to solve.

The aim of this work is to develop a solution method to relativistic quantum mechanical problems by solving the 
spin-$0$ Feshbach-Villars (FV0) equations. We adapt a solution method which worked extremely well in the
non-relativistic case. 
First, in Sec.\ II, we outline the FV0 formalism. Then in Sec.\ III we 
recapitulate the solution method for the non-relativistic problems with Coulomb-like potentials. 
In Sec.\ IV we present our approach to the FV0 equation. As an example, we choose an attractive Coulomb
plus short-range potential problem.
This also provides us a convenient test case since the Klein-Gordon hydrogen levels are well known 
from standard text books. Finally we summarize our findings and draw some conclusions. \\

\section{Feshbach-Villars equations for spin-$0$ particles}
\label{fv0-outline}

The Klein-Gordon equation with a  scalar potential $V$ is given by
\begin{equation}
\left(i\hbar \frac{\partial}{\partial t}  - V \right)^{2}  \Psi  = \left( c^{2} p^{2} + m^{2}c^{4} \right)  \Psi .
\end{equation}
In the FV0 formalism the wave function is split into two components
\begin{equation}
\Psi =  \phi + \chi,  \ \ \ \ \ \ \ 
\left( i \hbar \frac{\partial }{ \partial t} - V \right) \Psi  =  m c^{2} (\phi -\chi),
\end{equation}
and the components satisfy the coupled equations
\begin{eqnarray}
i\hbar \frac{\partial}{\partial t} \phi & = & \frac{p^{2}}{2m} (\phi + \chi) + ( mc^{2} +V) \phi  \label{fv1} \\
i\hbar \frac{\partial}{\partial t} \chi & = & -\frac{p^{2}}{2m} (\phi + \chi) - ( mc^{2} -V) \chi~.  \label{fv2}  
\end{eqnarray}
By introducing the two-component wave function
\begin{equation}
| \psi \rangle = \begin{pmatrix}  \phi \\ \chi \end{pmatrix} 
\end{equation}
and the Hamiltonian
\begin{eqnarray} \label{HFV0}
H_{FV0} 
&=& \begin{pmatrix}  1 & 1 \\ -1 & -1 \end{pmatrix} \frac{p^{2}}{2m} + 
	\begin{pmatrix}  1 & 0 \\  0 & -1 \end{pmatrix}  mc^{2} +  \begin{pmatrix}  1 & 0 \\ 0 & 1 \end{pmatrix}V \ \ \ \ \ \ \ \ 
\end{eqnarray}
we can write (\ref{fv1}) and (\ref{fv2}) into a form analogous to the time-dependent Schr\"odinger equation
\begin{equation}
i \hbar \frac{\partial}{\partial t} |\psi \rangle = H_{FV0} | \psi \rangle,
\end{equation}
or, for stationary states we may have
\begin{equation}
  H_{FV0} | \psi \rangle = E | \psi \rangle ~.
\end{equation}

This Hamiltonian looks like a usual coupled-channel Hamiltonian. 
However, there 
the channels are coupled by some short-range potential, 
while here the coupling is due the kinetic energy operator. 
As we are going to see below,  the matrix elements of the kinetic energy operator on a discrete basis
 representation behaves like $n$ for large $n$. This is a long-range coupling,
 it cannot be truncated to a finite basis representation.

\section{Solution method for non-relativistic problems }
\label{nonrel-outline}

We consider a Coulomb-like potential
\begin{equation}
 v_{l} = v^{C}+v^{(s)}_{l},
\end{equation}
where $l$ is the angular momentum, $v^{C}=Z e^{2}/r$ is the Coulomb potential and $v^{(s)}_{l}$ is some short-range potential.
The bound  states are the solutions of the homogeneous Lippmann-Schwinger equation
\begin{equation}
  |\psi \rangle = g^{C}(E) v_{l}^{(s)} |\psi \rangle,
\end{equation}
where
\begin{equation}
   g^{C}(E) = (E - p^{2}/2m - v^{C})^{-1}
\end{equation}
is the the Coulomb Green's operator.

If we approximate the short-range potential by a finite-rank operator, 
we can solve the equations without any further 
approximation, provided we can calculate the matrix elements of the Green's operator. 
In this respect the use of the Coulomb-Sturmian (CS) functions
\begin{equation}
\langle r | n \rangle = \left(\frac{n!}{(n+2l+1)!} \right)^{1/2} \mathrm{e}^{-br}(2br)^{l+1}L_{n}^{2l+1}(2br),
\end{equation}
where   $L$ is the Laguerre polynomial and $b$ is a parameter, 
is particularly advantageous. The CS functions form a basis. With 
$\langle r |  \widetilde{n} \rangle = \langle r | n \rangle /r
$
we have:
\begin{equation}
\langle \widetilde{n}  |n' \rangle = \delta_{n n'}
\end{equation}
and 
\begin{equation}
1 = \lim_{N\to\infty } \sum_{n=0}^{N} | n \rangle \langle \widetilde{n} | = \lim_{N\to\infty } \sum_{n=0}^{N} | \widetilde{n} \rangle \langle n | ~.
\end{equation}

The simple form of CS basis allow an exact and analytic calculations of the matrix
elements:
\begin{equation}
\langle n  |n' \rangle = 
	\begin{dcases}  (n+l+1)/b  & \text{ if  } n=n', \\
		-\sqrt{n'(n'+2l+1)}/(2b) & \text{ if  } n'=n+1, \\
		0  & \mbox{ otherwise,}
	\end{dcases}
\end{equation}
and
\begin{equation}
\langle n  | \frac{p^{2}}{2m }|n' \rangle = 
	\begin{dcases}  (n+l+1)\:b/(2m)  & \mbox{ if  } n=n', \\
		\sqrt{n'(n'+2l+1)}\:b/(4m) & \mbox{ if  } n'=n+1, \\
		0  & \mbox{ otherwise.}
	\end{dcases}
\end{equation}
Thus the operator $J =E-p^{2}/2m-v^{C}$ is an infinite tridiagonal matrix on the CS basis.
An important result of Refs.\ \cite{Konya:1997JMP,PRADemir2006} is that for infinite tridiagonal matrices the  
$N\times N$ part of the inverse, or the Green's matrix, is given by
\begin{equation}\label{greenc}
\underline{g}^{C} = (\underline{J} - \delta_{iN}\delta_{jN} J_{N,N+1} C_{N+1} J_{N+1,N} )^{-1},
\end{equation}
where $\underline{g}^{C}$ and  $\underline{J}$ are $N\times N$ matrices and $C$ is a continued fraction.
So, basically, $\underline{g}^{C}$ is almost the inverse of $\underline{J}$, only the right-down corner of
$\underline{J}$ is modified by a continued fraction. The continued fraction is
 constructed from the higher-index elements of $J$ and is 
defined by the recursion relation
\begin{equation}\label{cf}
C_{N+1}= (J_{N+1,N+1} - J_{N+1,N+2} C_{N+2}  J_{N+2,N+1})^{-1}.
\end{equation}
 
Over many years various finite-rank expansion methods for approximating $v^{(s)}$ have been used. 
In a recent work we proposed a simple, straightforward, 
yet very efficient approximation scheme   \cite{brown2013approximations}. 
It amounts of representing the short-range potential in a larger basis, invert the potential matrix, truncate to a smaller basis,
and then invert it back. This way we achieve a low-rank representation of the potential operator that  
contains the relevant information from the larger basis. Even a low-rank representation gives very good figures,
while higher-rank representations give extremely accurate results.

This method of solving problems of non-relativistic quantum mechanics through integral equations has been
extended further for solving more challenging problems like the three-body Faddeev integral equations with
Coulomb potentials \cite{PhysRevC.55.1080,PhysRevA.63.062721}.

\section{Relativistic spin-$0$ particles in Coulomb-like potentials} 

We consider a relativistic spin-$0$ particle in a Coulomb-like potential. Its FV0 Hamiltonian is given by
\begin{equation} \label{HFV00}
H_{FV0}  = H_{FV0}^{C} +  v_{FV0}^{(s)} 
\end{equation}
with
\begin{equation} \label{HFV00}
H_{FV0}^{C}  =   \begin{pmatrix}  1 & 1 \\ -1 & -1 \end{pmatrix}  \frac{p^{2}}{2m} +   
\begin{pmatrix}  1 & 0 \\ 0 & -1 \end{pmatrix} mc^{2} +  \begin{pmatrix}  1 & 0 \\ 0 & 1 \end{pmatrix} v^{C}
\end{equation}
and
\begin{equation} \label{HFV00}
v_{FV0}^{(s)}  = \begin{pmatrix}  1 & 0 \\ 0 & 1 \end{pmatrix} v_{l}^{(s)}.
\end{equation}
Analogous to the non-relativistic case, we can cast the eigenvalue problem of the two-component wave function 
into a Lippmann-Schwinger form
\begin{equation}\label{lsfv0}
  |\psi \rangle = g^{C}_{FV0}(E) v_{FV0}^{(s)} |\psi \rangle,
\end{equation}
where
\begin{equation}
   g^{C}_{FV0}(E) =  [ E - H_{FV0}^{C}   ]^{-1}.\ \ \ \ 
\end{equation}
 
Since $v_{FV0}^{(s)}$ is just a diagonal $2\times 2$ matrix with $v_{l}^{(s)}$ in the diagonal, we can approximate 
$v_{l}^{(s)}$ as before in the non-relativistic case.

In the CS basis the constant, the Coulomb potential and the kinetic energy operators are either diagonal or tridiagonal
infinite matrices. So, for
\begin{equation}
J_{FV0} =    \begin{pmatrix}  1 & 0 \\ 0 & 1 \end{pmatrix}E -  
 \begin{pmatrix}  1 & 1 \\ -1 & -1 \end{pmatrix}  \frac{p^{2}}{2m} -  
 \begin{pmatrix}  1 & 0 \\ 0 & -1 \end{pmatrix}  mc^{2} - 
\begin{pmatrix}  1 & 0 \\ 0 & 1 \end{pmatrix} v^{C},
\end{equation}
the infinite tridiagonal structure is superimposed by some $2\times 2$ matrix structures. 
In other words,  $J_{FV0}$ is 
infinite block-tridiagonal with $2\times 2$ matrix blocks. The same arguments which resulted in 
Eq.\ (\ref{greenc}) is applicable here and leads to a similar expression, only 
$J_{i,j}$'s should be understood as $2\times 2$ matrices and the continued fraction should be replaced by 
a continued fraction of $2\times 2$ matrices, i.e.\ the continued fraction becomes a matrix continued fraction.

In CS basis representation (\ref{lsfv0}) turns into a homogeneous algebraic equation
\begin{equation}
((\underline{g}_{FV0}^{C}(E))^{-1} - \underline{v}_{FV0}^{(s)})\underline{\psi}=0,
\end{equation}
which is solvable if the determinant is zero
\begin{equation}\label{det}
|(\underline{g}_{FV0}^{C}(E))^{-1} - \underline{v}_{FV0}^{(s)}|=0.
\end{equation}

\section{Numerical illustrations} 

In this work we use units such that $\hbar = 1$, $m=1$, $e^{2}=1$, and $c=137.03602$.
We consider first the hydrogen atom with $Z=-1$ and $l=0$.  The table below show the exact non-relativistic
and relativistic Klein-Gordon hydrogen energies, as well as the numerical results of the FV0 equations. 
We found that the numerical zeros of the determinant in (\ref{det}) coincide with the exact Klein-Gordon 
hydrogen eigenstates up to 10 digits, even for higher excited states, irrespective of $N$. 

Then we add a short-range potential
\begin{equation}
v^{(s)}= v_{0}\: {e^{-\alpha_{0} r}}/{r}
\end{equation}
with $v_{0}= 2$ and $\alpha_{0} = 2$, and investigate its effect on the bound states. The results are  shown
in the last column. We found that the method is accurate
enough to give account of the fine relativistic effects. We also experienced a rapid convergence in $N$, the number of states used in the expansion. The rate of convergence was similar that we observed in the 
non-relativistic case in Ref.\ \cite{brown2013approximations}. 

\begin{table}[htb]
\begin{center}
\tabcolsep=0.11cm
\begin{tabular}{ l | c | c |  c|c }
\hline \hline
n& Sch+$v^C$  & KG+$v^C$ & FV0+$v^C$ & FV0+$v^C$+$v^s$  \\ \hline
0 & -0.50000000 &  -0.50003329 & -0.50003329 &  -0.28203629  \\ \hline
 1& -0.12500000 &  -0.12500541 &  - 0.12500541 &  -0.09299089   \\ \hline
 2& -0.05555556 & -0.05555728 &- 0.05555728   &   -0.04546869   \\ \hline
 3& -0.03125000 & -0.03125075& -0.03125075     &   -0.02685280  \\ \hline
  4   & -0.02000000 &-0.02000039 & -0.02000039    &    -0.01770225  \\ \hline
  5   &-0.01388889  &-0.01388912 & -0.01388912      &   -0.01254053  \\ \hline
  6   &-0.01020408 &-0.01020423 & -0.01020423       &   -0.00934633 \\ \hline
  7   &-0.00781250  &-0.00781260 & -0.00781260     &    -0.00723344  \\ \hline
  8   &-0.00617284  &-0.00617291 & -0.00617291 &    -0.00576369   \\ \hline
  9   &-0.00500000  &-0.00500005 & -0.00500005   &  -0.00470027   \\ \hline
  10 &-0.00413223  &-0.00413227 & -0.00413227  &   -0.00390614   \\ \hline
\hline
\end{tabular}
\end{center}
\caption{Feshbach-Villars eigenstates of hydrogen atom (FV0$+v^{C}$), 
hydrogen plus short range potential (FV0$+v^{C}+v^{s}$) compared  to the 
exact Klein-Gordon (KG+$v^C$) and Schr\"odinger (Sch+$v^C$) results.}
\end{table}

\section{Summary and conclusions}

In this work we considered a relativistic spin-$0$ particle in a Coulomb plus short-range potential. 
We adopted the Feshbach-Villars formalism. In this formalism the Klein-Gordon wave function is split into two 
components and for the component we get a Schr\"odinger-like equation with a non-hermitian Hamiltonian.
We have rewritten the equations into a Lippmann-Schwinger form and represented the short-range part 
of the potential 
in CS basis by a finite matrix. We calculated the CS matrix elements Feshbach-Villars Coulomb Green's 
operator by a matrix continued fraction. This representation of the Green's operator is exact, thus even a 
small basis representation gives a faithful account of the whole spectra of the Klein-Gordon hydrogen atom.
The method gives very accurate results for the Coulomb plus short-range case as well.

This is another example of non-hermitian quantum mechanics \cite{bender}.
 The FV0  Hamiltonian is not hermitian, yet it has real eigenvalues. It  is
hermitian in a generalized sense \cite{davydov1976quantum,wachter2010relativistic}.
Interestingly, Feshbach-Villars equations can also be derived by requiring CPT symmetry for the
quantum mechanical equation \cite{ni2002advanced}.

The solution of the Feshbach-Villars equations pose a real challenge to
most of the solution methods due to the long-range--type kinetic energy in the coupling term.
However, we can wrap up the infinite block-tridiagonal matrix into the Green's matrix by using 
matrix continued fractions thus eliminating the main hurdle of working with Feshbach-Villars equations.

\newpage
\bibliographystyle{iopart-num}

\bibliography{inversion}

\end{document}